# Cross Spline Net and a Unified World[1]

Linwei Hu, Ye Jin Choi[2] and Vijayan N. Nair
Model Risk Management, Wells Fargo, Charlotte, NC, 28202, USA

**Abstract**

In today's machine learning world for tabular data, XGBoost and fully connected neural network (FCNN) are two most popular methods due to their good model performance and convenience to use. However, they are highly complicated, hard to interpret, and can be overfitted. In this paper, we propose a new modeling framework called cross spline net (CSN) that is based on a combination of spline transformation and cross-network (Wang et al. 2017, 2021). We will show CSN is as performant and convenient to use, and is less complicated, more interpretable and robust. Moreover, the CSN framework is flexible, as the spline layer can be configured differently to yield different models. With different choices of the spline layer, we can reproduce or approximate a set of non-neural network models, including linear and spline-based statistical models, tree, rule-fit, tree-ensembles (gradient boosting trees, random forest), oblique tree/forests, multi-variate adaptive regression spline (MARS), SVM with polynomial kernel, etc. Therefore, CSN provides a unified modeling framework that puts the above set of non-neural network models under the same neural network framework. By using scalable and powerful gradient descent algorithms available in neural network libraries, CSN avoids some pitfalls (such as being ad-hoc, greedy or non-scalable) in the case-specific optimization methods used in the above non-neural network models. We will use a special type of CSN, TreeNet, to illustrate our point. We will compare TreeNet with XGBoost and FCNN to show the benefits of TreeNet. We believe CSN will provide a flexible and convenient framework for practitioners to build performant, robust and more interpretable models.

## 1  Introduction

In today's machine learning world for tabular data, XGBoost and fully connected neural network (FCNN) are two most popular methods. For example, in Kaggle competitions, they have been shown to be the top two winning algorithms. They are also very convenient to use, without the need for careful feature engineering and the algorithm is scalable to large data.

However, the downside is they are black-box ML models. They are highly complicated, hard to interpret and can be overfitted (with large train-test performance gap) for small data. On the other hand, a lot of empirical studies suggest that for most tabular data, the structure is not too complicated. For example, Lou, et al. (2013) has shown that models with up to two-way interactions can achieve have very good performance. Therefore, in this paper, we propose a new modeling framework that is as performant, convenient and scalable as FCNN or XGBoost, and is also less complicated, more interpretable and robust. Our framework, cross spline net (CSN), is based on a combination of spline transformation and cross-network (Wang et al. 2017, 2021). In addition, CSN is flexible. With different choices of the spline layer, we can reproduce or approximate a set of non-neural network models, including linear and spline-based statistical models, tree, rule-fit, tree-ensembles (gradient boosting trees, random forest), oblique tree/forests, multi-variate adaptive regression spline (MARS), SVM with polynomial kernel, etc. Therefore, CSN unifies the above set of non-neural network models under the same neural network framework. This unification is not only elegant, but also gives them a "new life" by optimizing them under powerful and scalable gradient descent algorithms (like ADAM), instead of the original case-specific optimization methods.

---

[1] The views expressed in this paper are those of the authors and do not necessarily reflect those of Wells Fargo.
[2] Work done when this author was an intern at Wells Fargo.

We introduce the CSN framework in Section 2 and explain why it unifies and brings new life to a set of different non-neural network models. In this initial paper, we focus on one special type of CSN, TreeNet. We describe TreeNet in Section 3 and perform simulation studies in Section 4 to compare TreeNet with XGBoost and FCNN. The real data analysis is presented in Section 5. Finally, we summarize our work in Section 6.

## 2 Cross Spline Net Methodology

Our goal is to model the following:
$$F(x) = Polynomial(\Phi(x)), \tag{1}$$
where $\Phi(x)$ is a spline transformation function. This is a very powerful and general model form. The spline transformation adds more nonlinearity and flexibility, and the polynomial function accounts for variable interactions. In fact, statisticians have been working on such model form for a long time, with careful engineering on selecting the best spline transformation and interacting variables to keep the model parsimonious. Many non-neural network machine learning models also fall in this category, with much more complicated transformation and polynomial function. For regression tree or rule-fit, $F(x) = \sum_i a_i \left(\prod_{j \in Path(i)} I(x_j \lesseqgtr c_{ij})\right)$, where $a_i$ is the node value for leaf node $i$ in the tree, $x_j$ is one of the splitting variables leading to leaf node $i$, and $c_{ij}$ is the split point. So $\Phi(x) = \{I(x_j \lesseqgtr c_{ij})\}$. Tree ensemble (XGBoost, random forest) has the same form but much more complicated. For oblique tree/forests, each split is on a linear projection of the predictors (see for example, Breiman 2001; Rodriguez et al., 2006), hence $\Phi(x) = \{I(w_{ij}^T x \lesseqgtr c_{ij})\}$. For MARS (J. H. Friedman 1991), $\Phi(x) = \left\{(x_j - c_{ij})_+, (c_{ij} - x_j)_+\right\}$. For SVM, the spline transformation can be infinity dimensional and is not explicitly calculated; however, for $d$-th degree polynomial kernel, $F(x)$ is a polynomial function of degree $d$, hence $\Phi(x) = x$.

Cross spline net fits the class of models in Equation (1) in neural network and in a unified way. To do this, first, it is very easy to implement spline functions in neural network. The hinge function used in MARS is just RELU activation. The indicator function of $I(x \lesseqgtr c)$ can be well approximated by a sigmoid activation function $\sigma(\alpha + \beta x)$, with large, fixed value of $\beta$ (e.g., 20) and an appropriate value of $\alpha$. Figure 2-1 shows how the indicator function $I(x > 0.5)$ can be approximated with a sigmoid function. Similarly, the indicator function of $I(w^T x \lesseqgtr c)$ can be well approximated by $\sigma(\alpha + w^T x)$. Therefore, the only remaining piece is to implement polynomial function in neural network. This can be achieved with the cross layer proposed in Wang et al. (2017, 2021), where stacking $k$ cross layers will result in a $k + 1$ degree polynomial. One special CSN architecture with sigmoid basis is provided in Figure 3-1. To summarize, CSN is convenient to use, can be easily built using the existing TensorFlow or PyTorch library, and is flexible with different choices of spline transformations available.

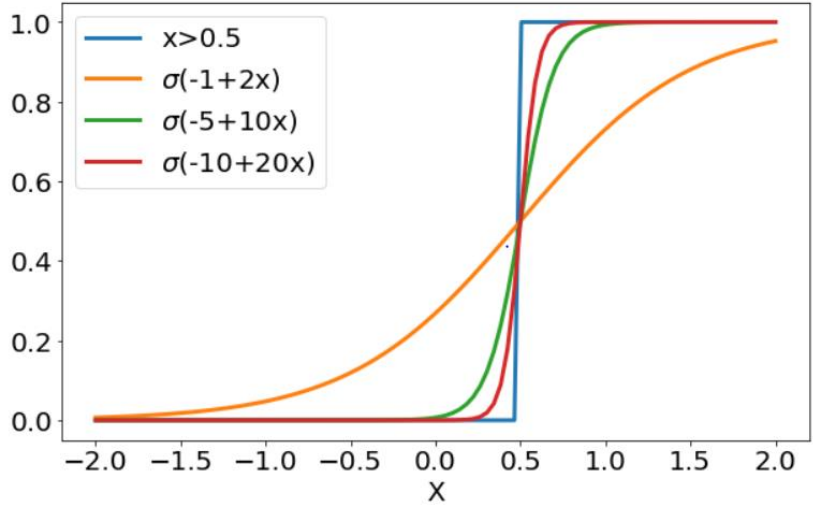

Figure 2-1. Indicator function and sigmoid function approximations

Note our proposed cross spline network has a key difference with the Deep & Cross network (DCN, DCN-v2) in Wang et al. (2017, 2021). Since the cross layer can only model polynomials, the authors supplemented it with a deep fully connected neural network after the cross layers, or alongside with the cross-layers to capture nonlinearity. However, we think it is better to handle nonlinearity before the cross layers with spline transformation. This is more intuitive from a statistical modeling perspective. More importantly, it is the key to CSN's better model interpretability and the unification of models in Equation (1). With FCNN entangled with cross-net, DCN and DCN-v2 are still black-box. On the other hand, CSN is less complicated and more interpretable when the polynomial function has a low degree. This can be easily controlled by the number of cross-layers. As empirical studies (e.g., Lou et al. 2013) have shown, tabular data usually has a low degree of interactions, so such constraint won't impact the model performance much. With simpler model structure, it is natural to have less overfitting and more robustness. Therefore, we have explained that CSN is flexible, convenient, performant, more interpretable and robust than black-box ML models.

Lastly, we will explain how the unification under CSN brings new life to the aforementioned non-neural network models. The first aspect comes from the optimization method. To see this, we point out the current optimization methods for these models are case-specific, can be somewhat greedy, ad-hoc or non-scalable to large data. For example, the implementation for tree, oblique tree, XGBoost, Random Forest, MARS, and so on are all non-trivial, and different from each other. Moreover, the optimization algorithm in those can be somewhat greedy or even ad-hoc. While they work well in general, it can be hard to find the global optimum sometimes. For example, tree is a greedy algorithm since it only considers one variable at a time during split. This means the splits are not optimized jointly. In particular, the root node is determined by main effects only, without taking into consideration the interaction effects. This makes it very hard to capture pure interaction effects, where the model does not have any main effects, i.e., $E(F(\boldsymbol{x})|x_j) = 0$. MARS employs a greedy forward selection algorithm. Oblique-tree/forest cannot optimize the projection direction $\boldsymbol{w}$ easily, hence they often try out some random projections. On the other hand, with powerful (and scalable) neural network optimization routines designed to find the global optimum (e.g., ADAM), CSN has a better chance in finding the best solution. This is especially true for oblique-tree/forests since projections are easily optimized in neural network, but also true for tree/tree-ensemble when we have pure interactions (see examples in Section 4).

Another aspect of "new life" is CSN can mimic but avoid some drawbacks of the non-neural network models. For example, trees use indicator function as basis, which is discontinuous. It requires multiple splits to model any continuous functions. All these mean it can be jumpy and is prone to overfitting. However, sigmoid basis can approximate both linear and indicator function (Figure 2-1), making it more versatile. CSN with sigmoid basis (called TreeNet, see Section 3) mimics tree but is smoother and less overfitting.

In the rest of this paper, we will focus on one specific type of CSN due to time constraint. The type of CSN we look at is called TreeNet.

## 3  A Special Case: TreeNet

TreeNet is aimed to mimic tree/tree-ensembles due to their popularity. It is CSN with sigmoid basis. As shown in Figure 2-1, sigmoid basis is versatile, it can approximate an indicator function when necessary, but can also be linear and smooth otherwise. The architecture of TreeNet is shown in Figure 3-1. Specifically, in the first layer after input, we create $m$ sigmoid basis for each input variable $x_j$, i.e., $\{\sigma(\alpha_{ij} + \beta_{ij} x_j), i = 1, \ldots, m, j = 1, \ldots, p\}$. There is a total of $mp$ basis. Not all bases will be important, so we insert a linear projection layer to reduce the dimension down to $d$. Next, we stack $k$ cross layers to model the polynomial function and finally output. The value $k$ determines the order of interactions in TreeNet. The maximum order of interactions can be modeled is $k + 1$. The hyper-parameter $k$ here plays a similar role as the depth of tree in tree-ensembles. Note in a FCNN, it is impossible to limit the order of interactions.

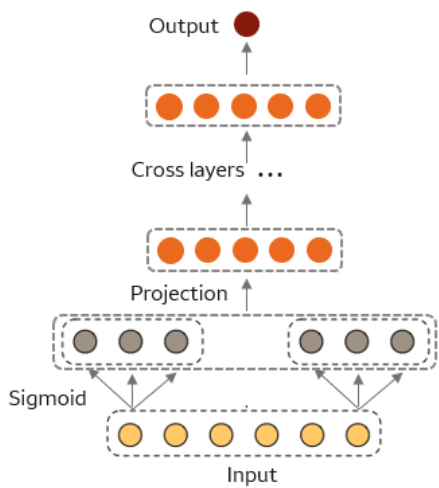

Figure 3-1. Architecture of TreeNet

### 3.1  Hyper-parameters

The hyper parameters $m, d, k$ have good default values. For $m$, since each sigmoid basis can approximate a local linear segment and the "knots" are optimized by the weights $\alpha$'s and $\beta$'s, we only need a small number of bases. Default $m = 5$ is sufficient since practical data don't have many waves. For $d$, it grows with the intrinsic dimension of the data. We experimented with $d = 20$ in our simulation studies and real data cases and it worked well. For $k$, the order of interaction is typically low for tabular data. Many studies suggest that a second order model ($k = 1$) can already fit very well to the data. Here we recommend $k = 2$ to fit up to 3-way interactions.

For the optimizer, by default, we use ADAM optimizer, with a learning rate of 0.02, batch size 1% of sample size, and learning rate decay of 0.995 per epoch. The best number of epochs is determined by early stopping. We call this TreeNet with the above default hyper-parameter settings TreeNet2. As we will see in Section 4 and 5, TreeNet2 works well in general and can be used as an off-the-shelf machine learning algorithm. However, if user wish to get the optimal model performance, these hyper-parameters can be tuned, and we provide a searching grid in Table 4-1.

Finally, it is known that all neural network models are "non-deterministic". Depending on the initialization, gradient descent can converge to global optimum or local optimum, yielding different models. In some rare cases, it may be trapped in a local optimum that's inferior. Tuning TreeNet or simply running the default hyper-parameter settings multiple times (with different random seed) will help the optimizer finding a good local optimum.

## 4  Simulation Study

In this section, we run some simulation studies to compare TreeNet with XGBoost and FCNN. We look at main effect only, two-way interaction and three-way interaction models. For each scenario, we have different flavors: continuous functional form, jumpy form (with indicator function), and pure interactions. The details are as follows:

- Main effect only
  - Main_cont: $f(x) = x_1 + 2x_2^2 + 2(1 + x_3)^{1/3} + 2x_4(x_4 > 0) + \sin(2\pi x_5) + e^{x_6}$, where all main effects are continuous.
  - Main_jump: $f(x) = x_1 + 2x_2^2 + 2(1 + x_3)^{1/3} + 2x_4(x_4 > 0) + I(x_5 > 0) + 2I(x_6 > 0.5)$, where $x_5, x_6$'s effects are discontinuous.
- Up to two-way interactions
  - 2way_cont: $f(x) = x_1 + 2x_2^2 + 2(1 + x_3)^{1/3} + 2x_4(x_4 > 0) + \sin(2\pi x_5) + e^{x_6} + 2x_1x_2 + 2\sin(\pi(x_3 + x_4)) + 2|x_5x_6|$, where all effects are continuous.
  - 2way_jump: $f(x) = x_1 + 2x_2^2 + 2(1 + x_3)^{1/3} + 2x_4(x_4 > 0) + I(x_5 > 0) + 2I(x_6 > 0.5) + 2(x_1 > 0)(x_2 > 0) + 2(x_3 > 0)(x_4 > 0) + 2x_5(x_6 > 0)$, where the interactions are discontinuous.
  - 2way_pure: $f(x) = 2x_1x_2 + \sin(\pi(x_3 + x_4)) + x_5\sin(\pi x_6)$, where all interactions are pure interaction (explained later)
- Up to three-way interactions
  - 3way_cont: $f(x) = x_1 + 2x_2^2 + 2(1 + x_3)^{1/3} + 2x_4(x_4 > 0) + \sin(2\pi x_5) + e^{x_6} + 2x_1x_2 + 2\sin(\pi(x_3 + x_4)) + 2|x_5x_6| + 3x_1e^{|x_2x_3|} + 3x_5\sin(\pi(x_4 + 1.5x_6))$, where all effects are continuous.
  - 3way_jump: $f(x) = x_1 + 2x_2^2 + 2(1 + x_3)^{1/3} + 2x_4(x_4 > 0) + I(x_5 > 0) + 2I(x_6 > 0.5) + 2(x_1 > 0)(x_2 > 0) + 2(x_3 > 0)(x_4 > 0) + 2x_5(x_6 > 0) + 3(x_1 > 0)(x_2 > 0.5)(x_3 < -0.5) + 3(x_4 > 0)(x_5 < -0.5)(x_6 < -0.5)$, where the interactions are discontinuous.

- 3way_pure: $f(x) = 2x_1x_2 + \sin(\pi(x_3 + x_4)) + x_5\sin(\pi x_6) + 2x_1x_2x_3 + 2x_4\sin(\pi(x_5 + x_6))$, where all interactions are pure interaction (explained later)

For each case, we simulate 30 predictors $x_1 \sim x_{30}$, following independent Uniform$(-1,1)$ distribution. Under such distribution, we have the conditional expectation $E(x_1x_2|x_1) = E(x_1x_2|x_2) = 0$, similarly for $\sin(\pi(x_3 + x_4))$ and $x_5\sin(\pi x_6)$. Hence the three interaction terms in 2way_pure are pure interactions with no main effects. This is an uncommon case, but we use it to show the drawback of tree/tree-ensembles in modeling pure interactions.

The response is simulated as $y = f(x) + \epsilon, \epsilon \sim N(0,1)$ for continuous case and $P(y = 1) = \frac{e^{\beta_0 + f(x)}}{1 + e^{\beta_0 + f(x)}}$ for binary case, where $\beta_0$ is selected such that the classes of 1's and 0's are balanced. A total of 10,000 observations are simulated for each combination of functional form and response type, divided into 70% training and 30% validation. A separate test set with 50,000 observations[3] is used to assess the test performance.

We fit five algorithms to our data: TreeNet, XGBoost, FCNN, TreeNet2 and XGBoost3. The first three algorithms are tuned to achieve optimal performance. Table 4-1 displays the values of the hyperparameters explored and a set of fixed parameters used throughout the model fitting process. The search spaces are carefully determined based on our thorough and extensive experiments. Search spaces for the hyperparameters of TreeNet are chosen based on their good default values discussed in the previous section. Note that the learning rate for TreeNet is larger because it has simpler structure than FCNN. This simpler structure does not require very small steps in gradient descent like FCNN does. Our experience shows that a learning rate of 0.01-0.02 works well. For all algorithms, we conduct a random search (with 20 trials) over the hyperparameters listed in the search space in Table 4-1. To prevent overfitting, we implement early stopping with a patience of 50 for all algorithms and L2 regularization with $\lambda = 1$ for XGBoost. For FCNN and TreeNet, we use the ADAM optimizer with a learning rate decay of 0.995 per epoch.

The last two algorithms (TreeNet2 and XGBoost3) come with fixed hyper-parameters. For TreeNet2, we fix number of cross layers at 2, number of bases at 5, dimension of projection layer at 20, learning rate at 0.02, batch size at 1% and select best number of epochs using early stopping. Since TreeNet2 is restricted to model up to three-way interactions, as a comparison, we also fitted XGBoost3, where max depth is fixed at 3, learning rate at 0.05, subsample at 0.7, L2 regularization at 1 and select best number of trees using early stopping.

Table 4-1. Hyper-parameter tuning settings

| Algorithm | Search Space | Fixed Hyperparameters |
|---|---|---|
| TreeNet | <ul><li>Learning rate: [0.01,0.02]</li><li>Batch size: [1%, 2%]</li><li>Number of cross layers: [0,1,2,3]</li><li>Number of bases: [3,5,7]</li><li>Dimension of projection layer: [10,20,30,40]</li></ul> | <ul><li>Decay=0.995/epoch</li><li>ADAM optimizer</li><li>Early stopping with 50 patience</li></ul> |
| FCNN | <ul><li>Layers: [20,10,5], [40,20,10], [60,30,15], [80,40,20], [100,50,25], [120,60,30,15],</li></ul> | |

---
[3] We used a larger sample size for test set so the test performance can be measured accurately.

| | | |
|---|---|---|
| | [100,50,25,12], [80,40,20,10], [10,20,40,20,10], [15,30,60,30,15]<br>• Learning rate: [0.0001,0.0005, 0.001,0.002,0.004,0.008, 0.01,0.015,0.02]<br>• Batch size: [1%, 2%, 4%] | |
| XGBoost | • Max depth: [1,2,3,4,5,6,7]<br>• Learning rate: [.01,.02,.03,.04,.05,.06,.07]<br>• Subsample: [.5, .7, 1] | • reg_lambda=1 (L2 regularization)<br>• Early stopping with 50 patience |

In Section 4.1 and Section 4.2, we show the results for the continuous and binary case, respectively. In Section 4.3, we increase the sample size from 10K to 50K and analyze the impact of sample size.

## 4.1 Continuous Response

The model performance for all five algorithms and eight cases are shown in Table 4-2. The testing MSEs are plotted in Figure 4-1 and the train-test performance gaps are plotted in Figure 4-2.

Table 4-2. Train MSE and Test MSE for 5 algorithms in 8 simulation studies

| | Train MSE | | | | | Test MSE | | | | |
|---|---|---|---|---|---|---|---|---|---|---|
| | TreeNet | FCNN | XGBoost | TreeNet2 | XGBoost3 | TreeNet | FCNN | XGBoost | TreeNet2 | XGBoost3 |
| **Main_cont** | 0.997 | 1.069 | 0.938 | 0.954 | 0.790 | **1.021** | 1.438 | 1.035 | 1.072 | 1.054 |
| **Main_jump** | 1.012 | 1.085 | 0.953 | 1.028 | 0.869 | 1.036 | 1.289 | **1.017** | 1.083 | 1.038 |
| **2way_cont** | 0.975 | 0.972 | 0.389 | 1.050 | 1.189 | **1.101** | 1.415 | 1.325 | 1.264 | 1.715 |
| **2way_jump** | 1.004 | 1.204 | 0.811 | 1.088 | 0.881 | 1.143 | 1.647 | **1.126** | 1.262 | 1.140 |
| **2way_pure** | 0.926 | 0.924 | 0.233 | 1.023 | 0.952 | **1.100** | 1.315 | 1.254 | 1.143 | 1.503 |
| **3way_cont** | 1.060 | 1.131 | 0.137 | 0.952 | 2.529 | **1.440** | 2.058 | 2.854 | 1.464 | 3.554 |
| **3way_jump** | 1.130 | 1.612 | 0.650 | 1.390 | 0.991 | 1.438 | 1.921 | **1.209** | 1.547 | 1.298 |
| **3way_pure** | 1.008 | 0.882 | 0.174 | 1.041 | 1.817 | **1.254** | 1.589 | 1.761 | 1.270 | 2.459 |

First, focusing on the testing performance, we see that TreeNet (represented by red circles in Figure 4-1) performs best in all cases except for the jump cases: main_jump, 2way_jump, and 3way_jump, where XGBoost is the best. This is expected because the underlying functions in these cases contain indicator functions, where XGBoost can perform very well with its natural splitting procedure. However, there is not much difference (~0.02) in the test MSE between TreeNet and XGBoost for the main_jump and 2way_jump cases. For the 3way_jump case, as we keep adding more jumpy interactions with higher order, the difference increases to 0.23 (1.209 vs 1.438). This indicates that TreeNet provides reasonable approximation to jumpy functions, and the loss in performance is small when the number of jumpy functions is small.

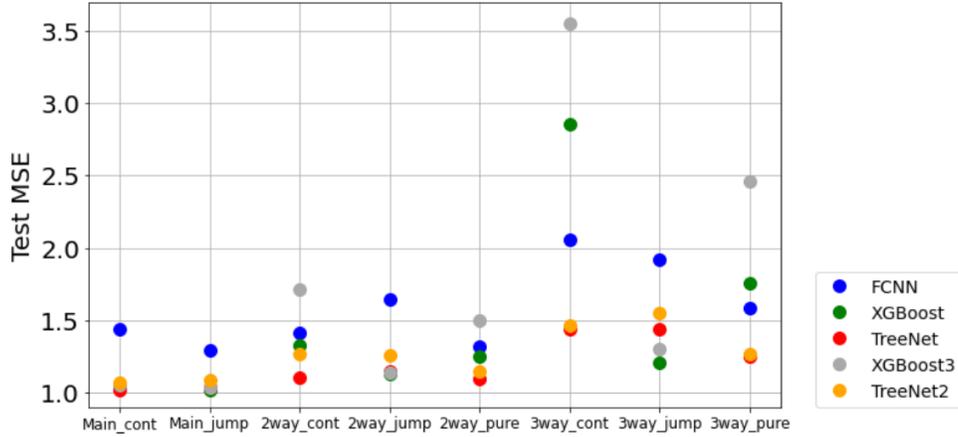

Figure 4-1. Test MSE for five algorithms in eight cases

XGBoost performs worse than TreeNet except for the three jump cases. It does especially poorly for 3way_cont (test MSE 2.854) and 3way_pure (test MSE 1.761). One of the reasons is XGBoost is less suitable for modeling smooth functions. The other reason, as explained in Section 2, pure interactions do not have any lower-order effects, so it is hard to capture for the greedy tree splitting algorithm which is driven by low-order effects at the top layers of the tree. For a $k$-way pure interaction effect, since there are no lower order effects, the top $k$ layers of the tree are constructed randomly. The only way a depth-$k$ tree can capture any signal is if any of its leaf nodes happens to split on all $k$ interacting variables. Given there are $p^k$ splitting variable combinations for a leaf node, it becomes exponentially harder when $k$ or $p$ is large. In the 3way_pure case, the XGBoost3 model with max depth = 3 has even worse performance (test MSE 2.459) than the tuned XGBoost, learning very little from the data. The tuned XGBoost has max depth 5, which makes it easier to capture 3-way interactions, but still significantly worse than TreeNet.

For FCNN (blue circles), it performs the worst in all cases except 3way_cont and 3way_pure, where it does better than XGBoost but worse than TreeNet. Certainly, it is not good at capturing jumpy patterns and does worse than TreeNet or XGBoost; but also, it does not do well in capturing some smooth functions. For example, it does not do well for main_cont, with test MSE significantly higher than TreeNet or XGBoost. The ICE plots in Figure 4-3 will show more details about this.

For TreeNet2 and XGBoost3, we can see TreeNet2 performs well, better than FCNN in all cases, and only slightly worse than the tuned TreeNet. It does worse than XGBoost3 for 2way_jump and 3way_jump, comparable or better for the other cases. It is much better especially for 3way_cont and 3way_pure.

Turning to the train-test performance gap (Figure 4-2), we see that TreeNet and TreeNet2 have the smallest gap. This confirms our theory that TreeNet overfits less than XGBoost since its basis functions are continuous, unlike XGBoost. As for FCNN, it inherently models high-dimensional interactions, making it more prone to overfitting than TreeNet and less suitable for low-dimensional interaction models. In conclusion, TreeNet has similar or better performance and less overfitting compared to the other models.

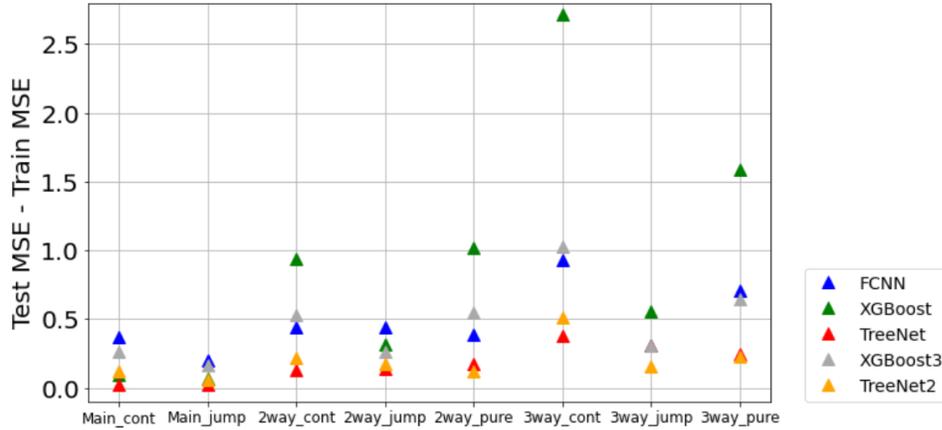

Figure 4-2. Difference between Test MSE and Train MSE

To further understand the advantages and drawbacks of each algorithm, we use ICE (Individual Conditional Expectation) plot to probe the fitted models and compare with the truth. ICE plot (see Goldstein et al., 2015) shows how the prediction changes for a selected observation, when the value of $x_j$ changes. Other diagnostics tools like partial dependence plot (PDP) can be used too, but we prefer ICE plot here because the PDP aggregates predictions, which won't be meaningful for the pure interaction cases. For each case, we randomly selected one sample and plotted the ICE curve against each variable. The ICE plot changes with respect to the sample we draw, but nevertheless, some patterns we observe are quite consistent.

Figure 4-3 presents the ICE curves for the tuned TreeNet, FCNN, and XGBoost in eight cases (results for TreeNet2 are similar to TreeNet, so we don't show it here. Similarly, we skip XGBoost3). FCNN fails to capture both wavy and discontinuous patterns. It produces continuous, piece-wise linear ICE plots due to RELU activation function, but it significantly deviates from the cubic root function ($x_3$ in Main_cont and Main_jump), the indicator function ($x_5$ and $x_6$ in Main_jump, and all features in 2way_jump and 3way_jump), and the sine function (being wavy). The ICE curves shows that FCNN only manages to roughly capture the general increasing trend for jumpy functions and the up and down trend for the sine functions. The failure to capture such complex pattern becomes severe as the form of underlying functions gets more complex, as seen in the 2way_pure, 3way_jump, and 3way_pure.

XGBoost demonstrates strong performance in approximating the true functions in main_cont and all three jump cases as expected. However, it doesn't do well in the other cases and performs poorly in the 3way_cont and 3way_pure cases. In particular, XGBoost produces jagged ICE plots, and it is challenging to capture pure interactions and interactions that are highly nonlinear or sinuous. From Figure 4-3, it is evident that true functions with respect to $x_3$ and $x_4$ are poorly captured by the XGBoost algorithm in 2way_cont, 2way_pure, 3way_cont, and 3way_pure, due to the term $\sin(\pi(x_3 + x_4))$ being highly nonlinear. Similar behaviors are observed for $x_6$ in 3way_cont due to $x_5 \sin(\pi(x_4 + 1.5x_6))$, and $x_5$ and $x_6$ in 3way_pure due to $x_4 \sin(\pi(x_5 + x_6))$. As the frequency of the sine function increases, it becomes even more challenging.

For all cases, TreeNet, in general, provides predictions that are most closely aligned with the true functions across all features except for 3way_jump (where XGBoost model has a better performance). It can well approximate various types of underlying functions, even the wavy sine functions that is challenging for FCNN and XGBoost. The response surface is smooth, unlike the jagged response surface by XGBoost; and it can approximate jump functions well when it is needed. Here with only 7000 training samples, we can see the step function for $x_5$, $x_6$ are already well approximated in main_jump case. Similar observation for the all the variables in the 2way_jump case. The approximation will get better with larger sample size.

To conclude, TreeNet is capable of explicitly fitting low-order interaction effects with the product between splines, and the splines and products are optimized in a global way (instead of greedily optimized in XGBoost). Consequently, it performs well in various scenarios, whereas FCNN and XGBoost are not robust to certain types of feature interactions or patterns.

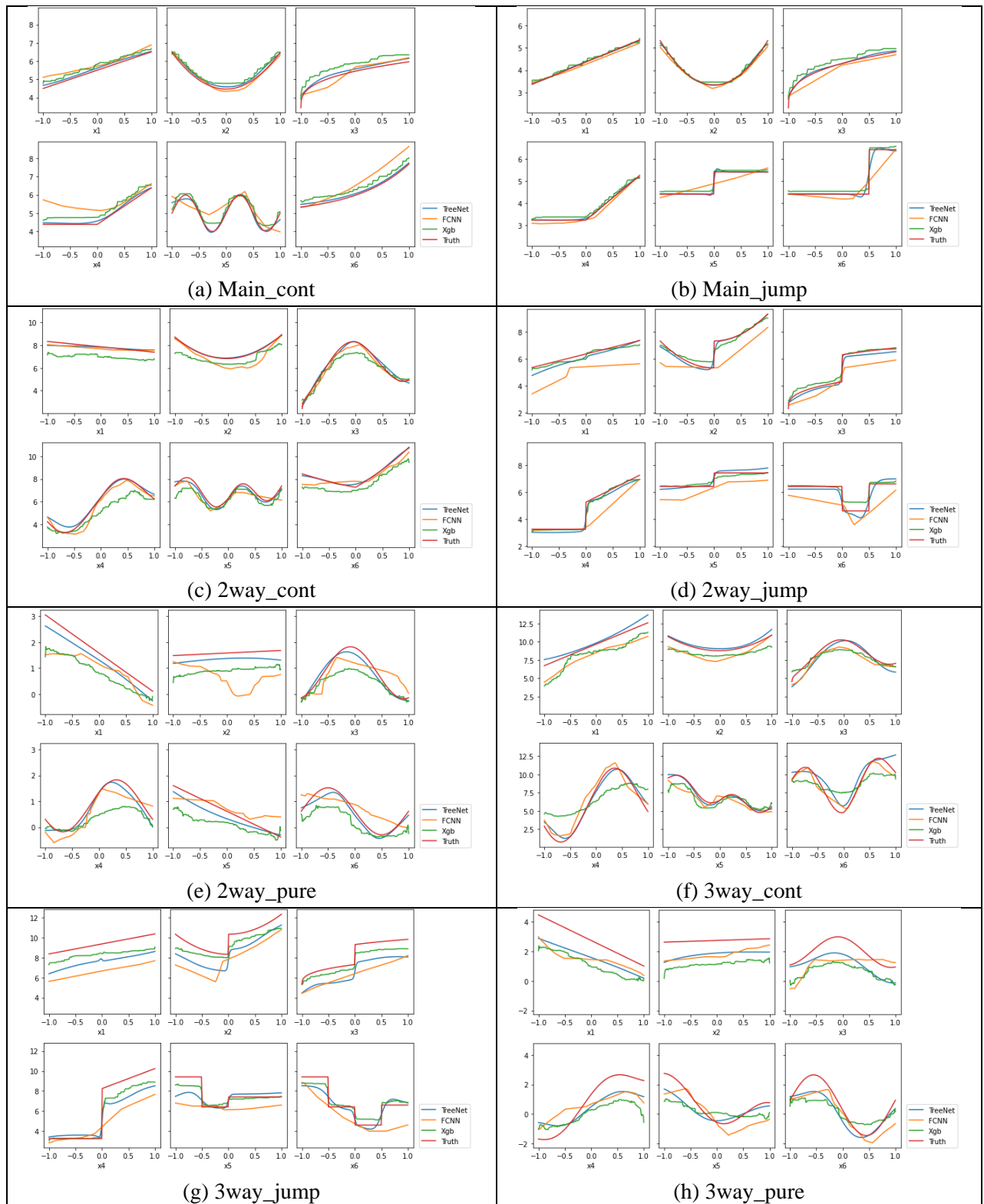

Figure 4-3. ICE plots with predictions from tuned TreeNet (blue), FCNN (orange), XGBoost (green) and true functions (red) in 8 simulation studies.

## 4.2 Binary Response

The results for binary response are qualitatively similar so we will briefly summarize it. The model performances are shown in Table 4-3 and the test AUCs are plotted in Figure 4-4. Again, we see that TreeNet (red circles) performs best in all cases except for the jump cases. Comparing TreeNet and XGBoost, ignoring small AUC difference of less than 0.01, TreeNet is better than XGBoost for 2way_pure and 3way_pure, and XGBoost is better for 3way_jmp and 2way_jump. The largest difference comes from 3way_pure, where TreeNet improves AUC by 0.067. For FCNN, it performs consistently worst in all cases, and the AUC is significantly worse in many cases. For TreeNet2, the performance is slightly worse than TreeNet (AUC difference ≤ 0.008) except for the 3way_pure case (AUC difference 0.034). It is better than XGBoost3 except for the jump cases and consistently better than FCNN.

Table 4-3. Train AUC and Test AUC for 5 algorithms in 8 simulation studies

|  | Train AUC | | | | | Test AUC | | | | |
| --- | --- | --- | --- | --- | --- | --- | --- | --- | --- | --- |
|  | TreeNet | FCNN | XGBoost | TreeNet2 | XGBoost3 | TreeNet | FCNN | XGBoost | TreeNet2 | XGBoost3 |
| **Main_cont** | 0.822 | 0.801 | 0.832 | 0.827 | 0.867 | **0.809** | 0.750 | 0.805 | 0.805 | 0.803 |
| **Main_jump** | 0.824 | 0.826 | 0.832 | 0.821 | 0.856 | 0.805 | 0.766 | **0.809** | 0.803 | 0.806 |
| **2way_cont** | 0.900 | 0.866 | 0.991 | 0.901 | 0.898 | **0.861** | 0.812 | 0.857 | 0.854 | 0.835 |
| **2way_jump** | 0.906 | 0.892 | 0.917 | 0.902 | 0.916 | 0.879 | 0.854 | **0.893** | 0.877 | 0.891 |
| **2way_pure** | 0.813 | 0.714 | 0.982 | 0.762 | 0.689 | **0.690** | 0.627 | 0.673 | 0.682 | 0.630 |
| **3way_cont** | 0.928 | 0.914 | 0.999 | 0.914 | 0.916 | **0.904** | 0.866 | 0.900 | 0.897 | 0.892 |
| **3way_jump** | 0.904 | 0.908 | 0.930 | 0.914 | 0.918 | 0.883 | 0.854 | **0.898** | 0.883 | 0.895 |
| **3way_pure** | 0.786 | 0.775 | 0.983 | 0.774 | 0.699 | **0.723** | 0.601 | 0.656 | 0.689 | 0.609 |

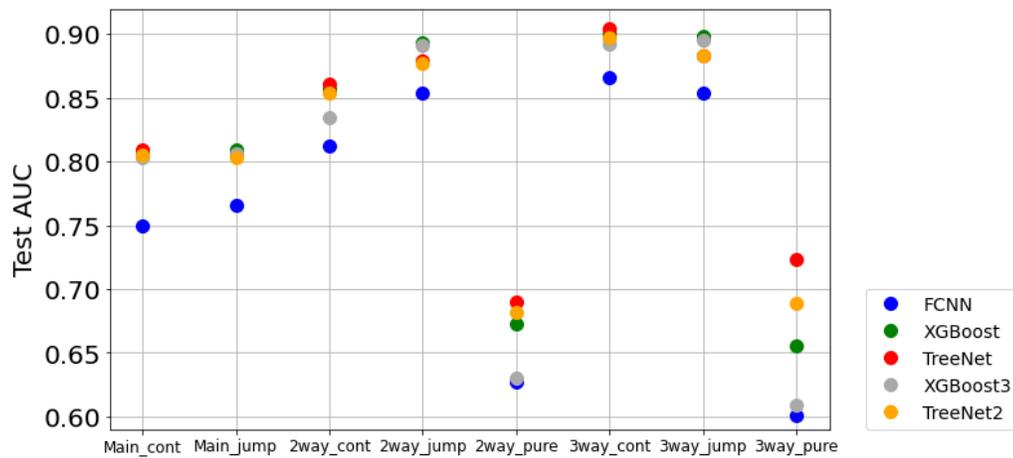

Figure 4-4. Test AUCs for five algorithms in eight cases

Like the continuous case, the overfitting gap between train AUC and test AUC (see Figure 4-5) is smallest for TreeNet, except for the 2way_pure case. For 2way_pure, FCNN and XGBoost3 have smaller gap because they haven't learned much from the data; the test AUC is much worse than TreeNet. The tuned XGBoost model has a worse but closer test AUC as TreeNet, but the overfitting gap is huge, indicating severe overfitting. For 3way_pure, not only the tuned XGBoost has a significantly worse AUC than TreeNet, but it also has huge overfitting gap. This supports our claim that fitting pure interactions are very hard for XGBoost model. Like TreeNet, TreeNet2 also has small overfitting gaps. To conclude, TreeNet/TreeNet2 has similar or better performance and less overfitting compared to the other models.

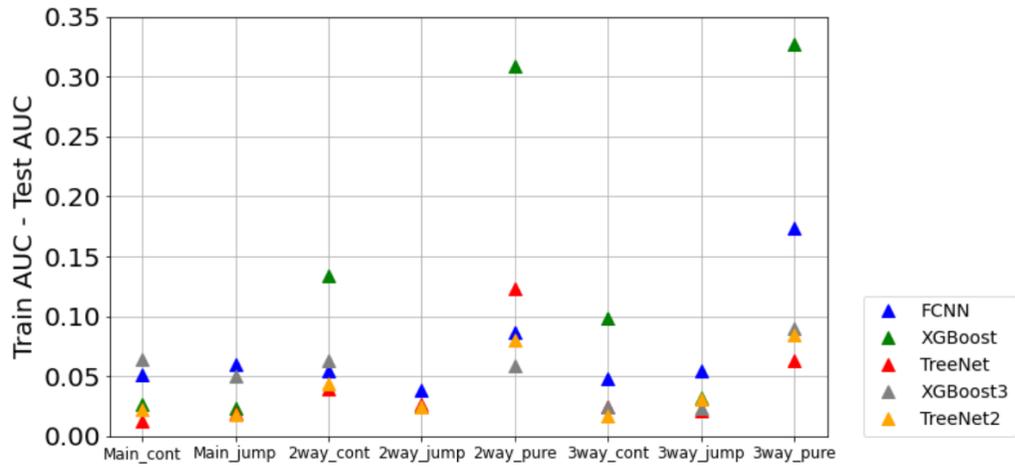

Figure 4-5. Difference between Test AUC and Train AUC

## 4.3 Impact of Sample Size

When sample size increases, all algorithms perform better and the previous worst algorithm, FCNN, sees the most improvement. As an illustration, we run the simulation study for 50K sample, for the 3way cases, both binary and continuous response. The model performance results are shown in Table 4-4, where the numbers in parenthesis are the performance numbers from the 10K sample. We can see first, TreeNet has close performance as XGBoost for the 3way_jump case, and still does significantly better than XGBoost in the other two cases. Secondly, FCNN's performance sees the largest improvement. For continuous case, the MSE greatly improves for the continuous cases (by 0.5-0.8), getting close to TreeNet and better than XGBoost for non-jump cases. Similar conclusion for binary case. TreeNet2 does well on all cases and only sees a tiny performance loss compared to TreeNet. XGBoost3 only does well for the jump cases and shows a significant performance loss compared to TreeNet or TreeNet2 for the other two cases.

The overfitting gaps become smaller as sample size increases for all algorithms. TreeNet still has smallest overfitting gap. The largest gap in MSE is less than 0.1 and the largest gap in AUC is 0.017. It means it almost does not overfit. FCNN has larger overfitting gap but smaller than XGBoost, and XGBoost has the largest overfitting gap.

Table 4-4. Model performance for 50K samples

| Continuous | Train MSE | | | | | Test MSE | | | | |
|---|---|---|---|---|---|---|---|---|---|---|
| | **TreeNet** | **FCNN** | **XGBoost** | **TreeNet2** | **XGBoost3** | **TreeNet** | **FCNN** | **XGBoost** | **TreeNet2** | **XGBoost3** |
| **3way_cont** | 1.017 | 0.994 | 0.497 | 0.985 | 2.321 | **1.108** | 1.222 | 1.966 | 1.116 | 2.887 |
| | (1.060) | (1.131) | (0.137) | (0.952) | (2.529) | (1.440) | (2.058) | (2.854) | (1.464) | (3.554) |
| **3way_jump** | 1.077 | 1.056 | 0.833 | 1.027 | 1.070 | 1.137 | 1.151 | **1.087** | 1.199 | 1.187 |
| | (1.130) | (1.612) | (0.650) | (1.390) | (0.991) | (1.438) | (1.921) | (1.209) | (1.547) | (1.298) |
| **3way_pure** | 0.994 | 1.018 | 0.388 | 1.019 | 1.418 | **1.082** | 1.094 | 1.306 | 1.088 | 1.858 |
| | (1.008) | (0.882) | (0.174) | (1.041) | (1.817) | (1.254) | (1.589) | (1.761) | (1.270) | (2.459) |
| **Binary** | Train AUC | | | | | Test AUC | | | | |

|  | TreeNet | FCNN | XGBoost | TreeNet2 | XGBoost3 | TreeNet | FCNN | XGBoost | TreeNet2 | XGBoost3 |
|---|---|---|---|---|---|---|---|---|---|---|
| **3way_cont** | 0.946 | 0.940 | 0.965 | 0.945 | 0.923 | **0.940** | 0.926 | 0.919 | 0.937 | 0.907 |
|  | (0.928) | (0.914) | (0.999) | (0.914) | (0.916) | (0.904) | (0.866) | (0.900) | (0.897) | (0.892) |
| **3way_jump** | 0.907 | 0.899 | 0.919 | 0.903 | 0.913 | 0.898 | 0.879 | **0.905** | 0.896 | 0.904 |
|  | (0.904) | (0.908) | (0.930) | (0.914) | (0.918) | (0.883) | (0.854) | (0.898) | (0.883) | (0.895) |
| **3way_pure** | 0.795 | 0.796 | 0.851 | 0.800 | 0.726 | **0.778** | 0.767 | 0.738 | 0.772 | 0.670 |
|  | (0.786) | (0.775) | (0.983) | (0.774) | (0.699) | (0.723) | (0.601) | (0.656) | (0.689) | (0.609) |

With improved model performance, the ICE plot is also better aligned with the truth. Figure 4-6 shows the ICE plots for the 3way cases for continuous response. We compare the ICE plots for the 50K sample models with 10K sample models (as seen in Figure 4-3). We can see better alignment overall, especially for FCNN. TreeNet is the best, being very close to the truth. FCNN sees the biggest improvement, especially for the jump functions. XGBoost still shows some gap with the truth for the non-jump cases. This is consistent with the worse model performance we observe.

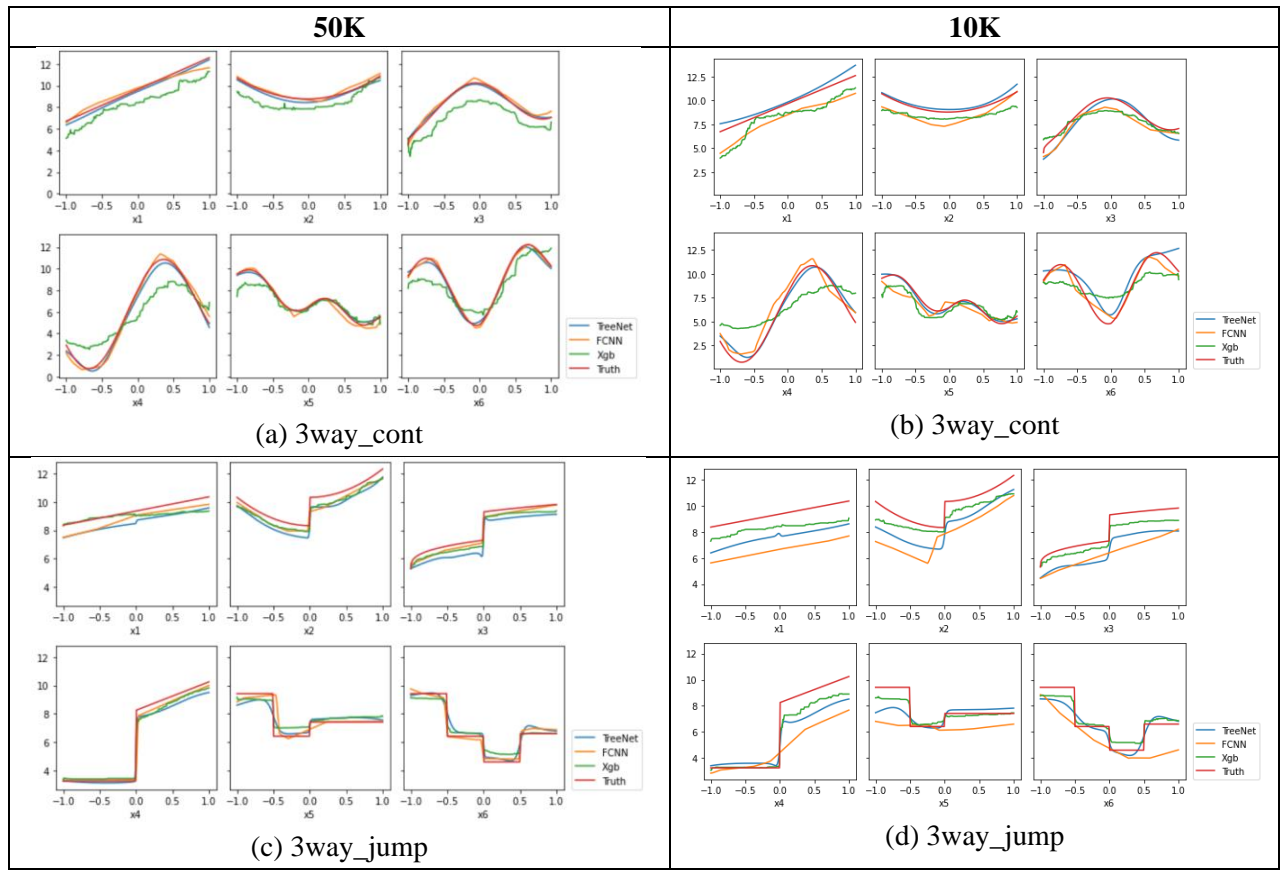

(a) 3way_cont     (b) 3way_cont

(c) 3way_jump     (d) 3way_jump

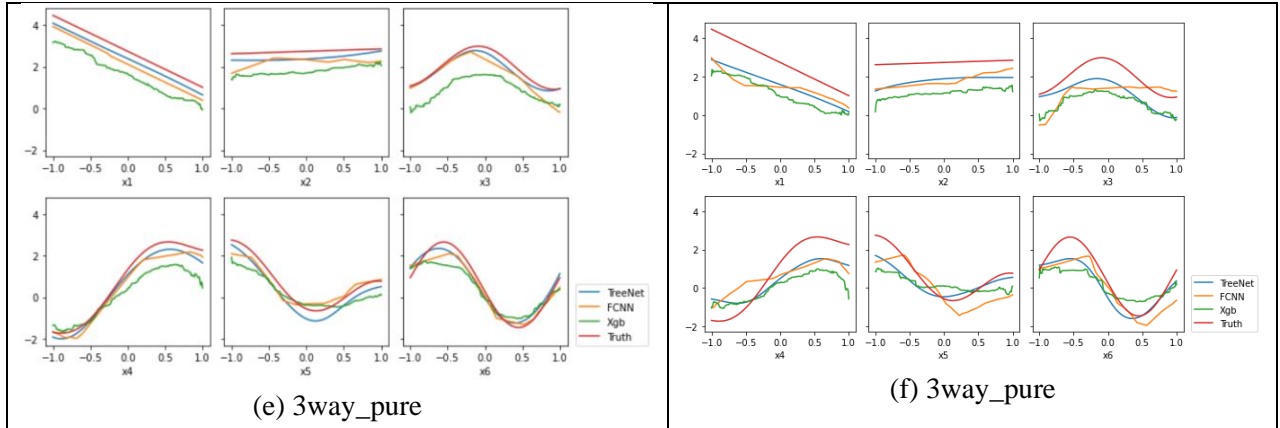

(e) 3way_pure          (f) 3way_pure

Figure 4-6. ICE plots for 50K (left) and 10K samples (right)

## 5 Real Data Analysis

In this section, we demonstrate the effectiveness of TreeNet by applying it to two real datasets. We compare TreeNet's performance with XGBoost and FCNN. The description of the two datasets are as follows:

- **Home Lending**: This dataset contains Wells Fargo's residential mortgage account information. The goal is to predict which home mortgage accounts are likely to be in "trouble" (1 if trouble and 0 otherwise) in the future. Trouble is defined as any of the following events: bankruptcy, short sale, 180 or more days of delinquency in payments, etc. There are over 50 predictors, including macro-economic variables at the future prediction time (unemployment rate, house price index, and so on), static loan characteristic variables at the origination time (loan types, property types, etc), and dynamic loan characteristic variales at the current time, or snapshot time (fico, delinquency status, loan-to-value ratio, etc). We selected a small sample with 1 million observations from one of the portfolios. Some key variables are listed in Table A-1 in the Appendix.

- **Bike Sharing**: This is a public dataset hosted on UCI machine learning repository. It contains hourly bike rental records from 2011 to 2012 in Washington DC area. The goal is to predict the hourly bike rental counts. For count data, it is typical to apply the log transformation to stabilize the variance, so we model the log-count instead. There are 11 predictors, including time information (hour, day of week, month, etc.) and weather information (temperature, humidity, wind speed, etc.). The total sample size is around 17K.

For each dataset, we divide into training (50%), validation (25%) and testing (25%). We use the training and validation set to tune TreeNet, FCNN and XGBoost, and use the testing set to assess the performance. As in Section 4, we also fit TreeNet2 and XGBoost3. The model performances are shown in Table 5-1 and Table 5-2. For home lending data, all algorithms have very close test AUC. The difference is only 0.001. For bike sharing data, XGBoost is the best, slightly outperform TreeNet (test MSE is 9% larger), followed by FCNN. XGBoost3 is very close to XGBoost, and TreeNet2 is very close to TreeNet, indicating there is little high order effect. In terms of overfitting gap, again, XGBoost has largest gap, and TreeNet has smallest gap.

Table 5-1. AUC for home lending data

|       | TreeNet | FCNN  | XGBoost   | TreeNet2 | XGBoost3 |
|-------|---------|-------|-----------|----------|----------|
| Train | 0.865   | 0.865 | 0.907     | 0.867    | 0.888    |
| Test  | 0.857   | 0.857 | **0.858** | 0.857    | 0.858    |

Table 5-2. MSE for bike sharing data

|       | TreeNet | FCNN  | XGBoost   | TreeNet2 | XGBoost3 |
|-------|---------|-------|-----------|----------|----------|
| Train | 0.099   | 0.073 | 0.049     | 0.094    | 0.083    |
| Test  | 0.106   | 0.120 | **0.097** | 0.108    | 0.100    |

Next, we compare the three main algorithms, tuned TreeNet, FCNN and XGBoost using model diagnostic tools. First, we look at permutation based variable importance. Figure 5-1 shows the variable importance for top 20 most important variables for home lending data. We can see they are quite consistent in the top-ranking variables. XGBoost and TreeNet share 9 common variables out of the top 10, and FCNN share 8 common variables out of top 10 with XGBoost or TreeNet.

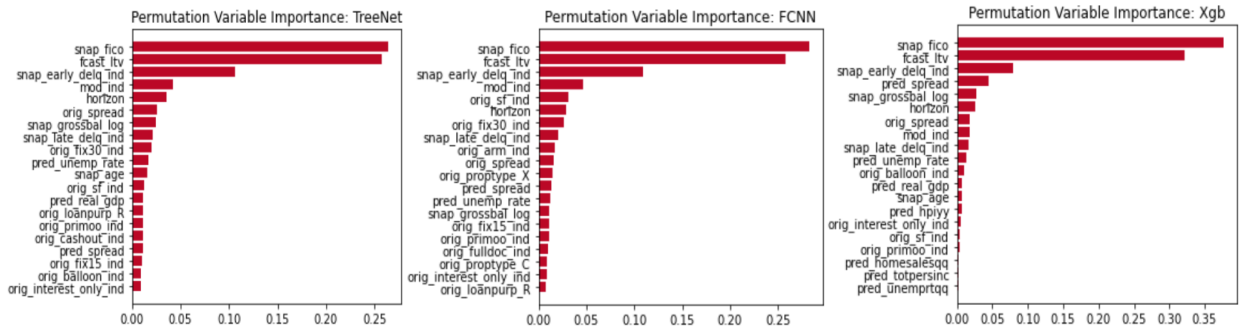

Figure 5-1. Variable importance for top 20 variables for home lending data

For the top three variables, we plot the 1d PDP in Figure 5-2. FCNN and TreeNet are closely aligned with each other, and they are smooth compared to XGBoost. XGBoost has some kinks, and on sparse data region (tail or when early delinquency indicator equals 1), we see some difference with TreeNet or FCNN.

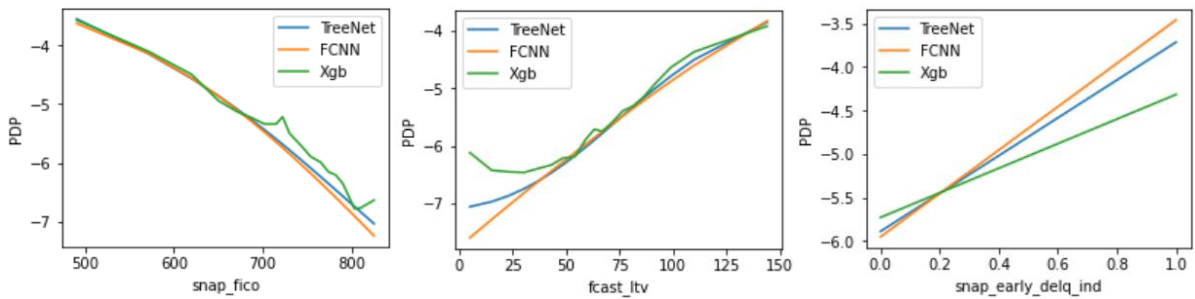

Figure 5-2. 1d PDP for top three variables for home lending data

We also look at two-way interaction effects. Figure 5-3 shows the 2d PDP for the top 2 interaction effects with largest H-statistics (see Friedman and Popescu 2008). The patterns are quite consistent. We can see the effect of horizon changes direction, depending on early delinquency status. On the other hand, the effect of fico always has a decreasing trend, but it becomes less sensitive with larger loan-to-value ratio. There exist also some small differences. For example, XGBoost has more kinks in the PDP (like we see in the 1d PDP), and FCNN shows stronger linearity effect (due to the RELU activation function).

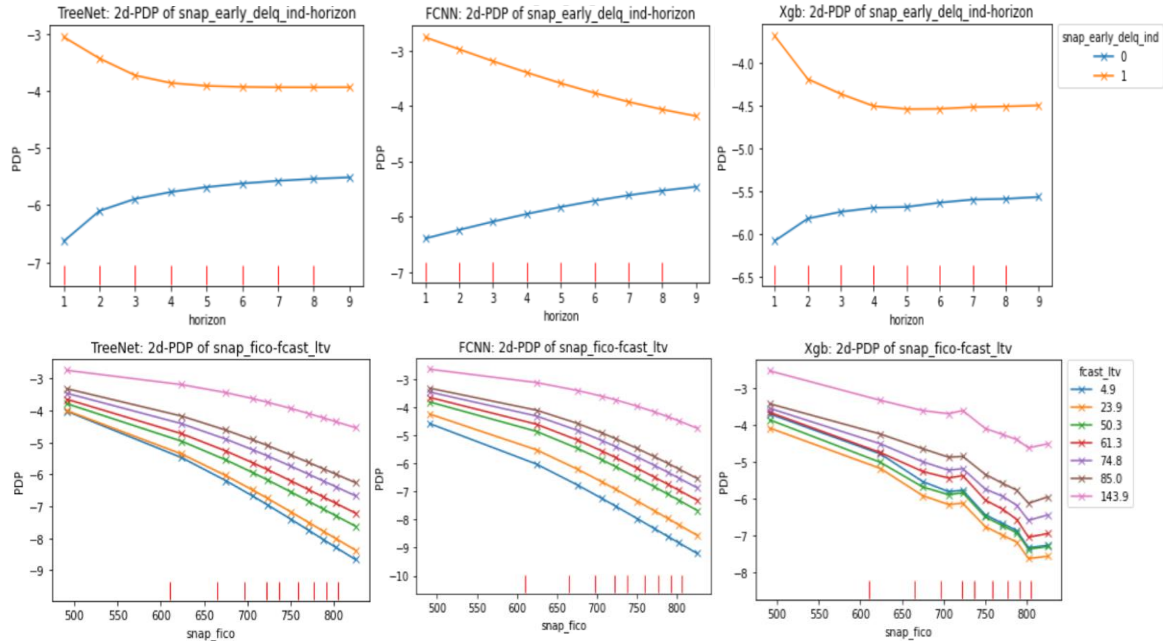

Figure 5-3. 2D PDP for top two interaction pairs for home lending data

For bike sharing data, the variable importance scores for the three tuned algorithms are shown in Figure 5-4. The top three most important variables are the same. In particular, hour is predominantly the most important variable, accounting for a majority of the importance scores.

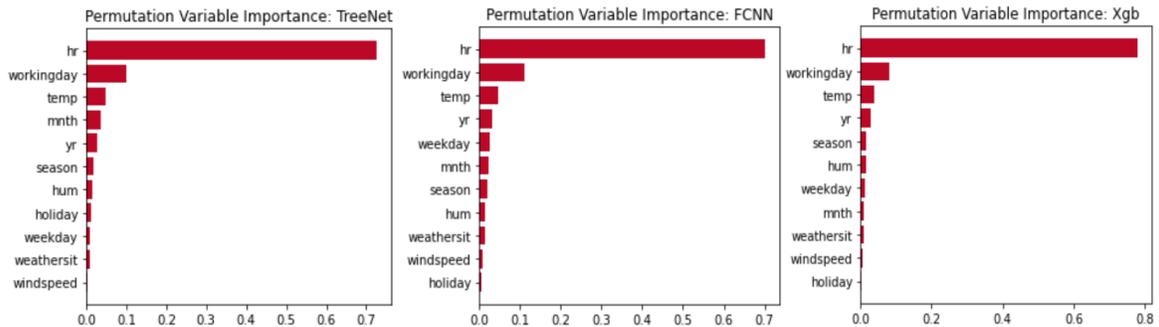

Figure 5-4. Variable importance for bike sharing data

The 1d PDPs for the top three variables are shown in Figure 5-5. For hour, the pattern is almost identical for all three algorithms. For the other two variables, TreeNet is close to XGBoost, while FCNN shows some differences. This might be due to TreeNet and XGBoost have closer model performance, and both are slightly better than FCNN.

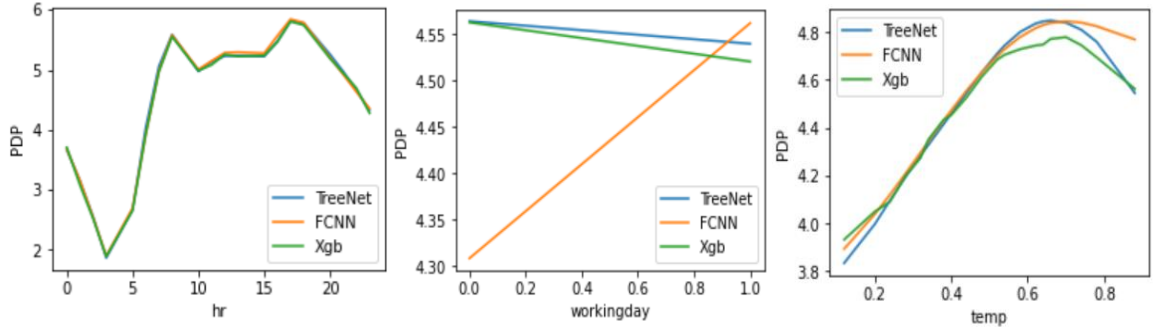

Figure 5-5. 1d PDP for top three variables for bike sharing data

Finally, we look at the two-way interaction effects. Hour and working day have the largest interaction effect in terms of H-statistics, and much bigger than the other interactions. Figure 5-6 shows the 2d PDP for this interaction. The pattern is quite consistent across models. For non-working day, the bike rental peaks around noon, whereas for working day, the bike rental peaks at the morning and afternoon rush hours.

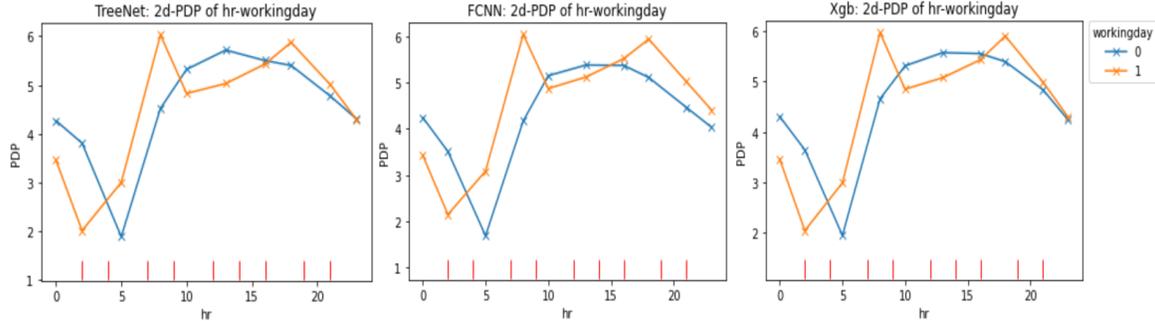

Figure 5-6. 2D PDP for the most important interaction for bike sharing data

# 6   Summary

In this paper, we proposed a new machine learning framework, cross-spline net (CSN), and used one special case, TreeNet, as an illustration. While we focused on TreeNet in our study, the framework itself is more important and much broader. The benefits of the CSN framework are enormous, including

- Flexibility: With a simple change of the spline basis, CSN can reproduce or approximate a variety of different models.
- Convenience: CSN can be easily implemented with the existing neural network modules and optimization routines in TensorFlow or PyTorch.
- Performant: CSN is optimized in a global way using advanced neural network optimization techniques (e.g., ADAM), so the performance is good. Plus, small number of cross-layers suffices to model tabular datasets since the order of interaction is usually low.
- Better interpretability: With small number of cross-layers, the model is simpler compared to black-box ML models and can be expanded into a low-degree polynomial function, hence more interpretable.
- More Robustness: As the model is simpler, it is less overfitting and more robust.
- Unification: We have unified a variety of non-neural network statistical/machine learning models under the same neural network framework, significantly enriches what neural network models can do. It also brings new life to these non-neural network models, by replacing the original ad-hoc or greedy optimization routines with scalable and power neural network gradient descent algorithms.

We believe that this framework will provide a new approach to build performant, robust, and more interpretable models. In the future, we will study the interpretability of CSN, model pruning, and other special types of CSN.

**Appendix**

Table A-1. Variables for home lending data

| Variable | Definition |
| --- | --- |
| horizon | prediction horizon (difference between prediction time and snapshot time) in quarters |
| snap_fico | credit score (FICO score) at snapshot time |
| orig_fico | credit score (FICO score) at loan origination |
| snap_ltv | loan to value (ltv) ratio at snapshot time |
| fcast_ltv | loan to value (ltv) ratio forecasted at prediction time |
| orig_ltv | loan to value ratio at origination |
| orig_cltv | combined ltv at origination |

| snap_early_delq_ind | early delinquency (no min payments for a few months) indicator: 1 means loan has early delinquency status at snapshot time; 0 means loan is current or has late delinquency status. 7.7% observations are early delinquent. |
|---|---|
| snap_late_delq_ind | late delinquency indicator (loan is delinquent for longer time, close to default) indicator: 1 means loan has late delinquency status at snapshot time; 0 means loan is current or has early delinquency status. Only 0.2% observations are late delinquent. |
| pred_loan_age | age of loan (in months) at prediction time |
| snap_gross_bal | gross loan balance at snapshot time |
| orig_loan_amt | total loan amount at origination time |
| pred_spread | spread (difference between note rate and market mortgage rate) at prediction time |
| orig_spread | spread at origination time |
| orig_arm_ind | Indicator: 1 if loan is adjustable-rate mortgage (ARM); 0 otherwise |
| pred_mod_ind | modification indicator: 1 means prediction time before 2007Q2 (financial crisis); 0 if after |
| pred_unemp_rate | unemployment rate at prediction time |
| pred_hpi | house price index (hpi) at prediction time |
| orig_hpi | hpi at origination time |
| pred_home_sales | home sales data at prediction time |
| pred_rgdp | real GDP at prediction time |
| pred_totpersinc_yy | total personal income growth (from year before prediction to year of prediction time) |